\documentclass[prd,showpacs,twocolumn]{revtex4}

\newcommand{\beq}{\begin{equation}}
\newcommand{\eeq}{\end{equation}}
\newcommand{\beqar}{\begin{eqnarray}}
\newcommand{\eeqar}{\end{eqnarray}}

\ifx\undefined\fiverm
     \font\fiverm=cmr5
\fi

\input{prepictex}
\input{pictex}
\input{postpictex}
\begin{document}
\author{Tarun Biswas}
\title{Through the Black Hole -- On Not Breaking Time Reversal Symmetry.}
\email{biswast@newpaltz.edu}
\affiliation{State University of New York at New Paltz, \\ New Paltz,  NY 12561, USA.}
\date{\today}
\begin{abstract}
It is well-known that a particle falling into a black hole will definitely reach the center
in finite proper time if it enters the sphere of radius $3r_{s}/2$ where $r_{s}$ is the
Schwarzschild radius. It is usually assumed that once the particle reaches the central singularity,
it stops. Here it shall be shown that there are no theoretical reasons for this assumption.
In fact, due to the time-reversal symmetry of the equation of motion, it is more ``natural''
to assume that the particle will travel through the singularity and come out on the other side. Of
course, it is not possible to compute the trajectory of the particle at the singularity itself. However,
one may compute the trajectory just before entry and just after exit. The continuity of the two
pieces at the singularity is maintained through energy and angular momentum conservation conditions.
The results of such computations are shown here. Also, for the particle to come to a stop at the center,
there must exist nonconservative forces at that point. Such forces being unknown both
theoretically and experimentally, it is prudent to disregard them.
\end{abstract}
\pacs{04.70.-s, 97.60.Lf}
\maketitle
\section{Introduction}
The three discrete symmetries, parity ($P$), charge conjugation ($C$) and time reversal ($T$), have long been
subjects of study. On the surface, it seems reasonable for each one of them to be a universal symmetry. Hence, when parity
violation was experimentally verified, it was a surprise\cite{leeyang,wu,leder}. Since then, the weaker condition
of the combination, $PCT$, is generally accepted as a universal symmetry\cite{streat}.

So, a violation of $T$ symmetry alone is not a physical impossibility. However, in the case of gravity, there are no
experimental or theoretical reasons to assume such a violation. Nonetheless, it is generally assumed 
that a particle entering the central singularity
of a black hole will stop right there. The equation of motion, being $T$ symmetric, requires the particle to 
exit the center just as it entered. So, to make the particle stop at the center, some physical mechanism
needs to be introduced by hand such that $T$ symmetry is broken at the origin. As the stopping of the particle
is like an inelastic collision, such a mechanism would have to be a nonconservative force. It is sometimes 
argued that the singularity itself provides this mechanism as it produces
an infinite effective force which the particle cannot escape. However, at the singularity, the particle also
possesses infinite effective kinetic energy. This should allow it to escape the infinite force unless
some other nonconservative force depletes its kinetic energy.

Here it shall be shown that $T$ symmetry violation is unnecessary for particle trajectories through a black hole.
Indeed, it is physically more ``natural'' to maintain $T$ symmetry and allow particle trajectories to run through
the center of a black hole. Actual trajectories through the singularity are computed using conditions derived
from angular momentum and energy conservation.

\section{Coordinate Systems and Time Reversal}
The Schwarzschild line element in standard spherical polar coordinates $(t,r,\theta,\phi)$ is given as
\beq
	d\tau^{2}=\left(1-\frac{r_{s}}{r}\right)dt^{2}-\left(1-\frac{r_{s}}{r}\right)^{-1}dr^{2}
	-r^{2}d\Omega^{2},
\eeq
where
\beq
	d\Omega^{2}=d\theta^{2}+\sin^{2}\theta d\phi^{2},
\eeq
the speed of light $c=1$, the Schwarzschild radius $r_{s}=2GM/c^{2}$, $G$ is the universal gravitational constant
and $M$ is the mass of the source. The equation of motion (geodesic) of a point particle\footnote{Only non-zero mass
particles are considered here. The equations of motion for zero mass particles are different. But they are also $T$
symmetric.} in this metric is found to be as follows\cite{rindler, night}.
\beqar
\frac{d^{2}u}{d\phi^{2}}+u & = & \frac{r_{s}}{2h^{2}}+\frac{3r_{s}u^{2}}{2}, \label{eqmotion} \\
r^{2}\frac{d\phi}{d\tau} & = & h, \label{eqangmom}
\eeqar
where $\theta=\pi/2$ gives the plane of the orbit, $u=1/r$, $\tau$ is the proper time
and $h$ is the conserved angular momentum for a particle of unit mass. For the special case of $h=0$, the motion is radial.
This makes equation~\ref{eqmotion} meaningless and it needs to be replaced by the following.
\beq
\frac{d^{2}r}{d\tau^{2}}+\frac{r_{s}}{2r^{2}}=0. \label{eqradial}
\eeq
The time reversal operation $T$ is given by
$\tau\rightarrow -\tau$ and $h\rightarrow -h$. So, the equation of motion is covariant under $T$. Consequently, particle
trajectories are expected to be $T$ symmetric. However, trajectories that stop at $r=0$ clearly do not possess this 
$T$ symmetry. Hence, for them, $T$ symmetry must be broken explicitly. Gravity of a 
black hole cannot do this by itself. The $T$ symmetry breaking agent must be some sort of nonconservative 
(nongravitational) force introduced by hand.

At this juncture, it is interesting to explore the connection of the Eddington-Finkelstein coordinates to $T$ symmetry.
For the ingoing Eddington-Finkelstein coordinates\cite{rindler,misner}, the $t$ coordinate is replaced by $V$ which is defined as follows.
\beq
	V=t+r+r_{s}\ln|r/r_{s}-1|.
\eeq
Using this, the Schwarzschild line element takes the following form.
\beq
	d\tau^{2}=(1-r_{s}/r)dV^{2}-2dVdr-r^{2}d\Omega^{2}.
\eeq
For the outgoing Eddington-Finkelstein coordinates, the $t$ coordinate is replaced by $U$ which is defined as follows.
\beq
	U=t-r-r_{s}\ln|r/r_{s}-1|.
\eeq
Using this, the Schwarzschild line element takes the following form.
\beq
  d\tau^{2}=(1-r_{s}/r)dU^{2}+2dUdr-r^{2}d\Omega^{2}.
\eeq
The introduction of these coordinates may be seen as a formal mechanism for breaking $T$ symmetry\cite{bis}.
The two versions of the Eddington-Finkelstein coordinates (ingoing and outgoing) effectively separate the ingoing
and outgoing geodesics of a black hole. So, discarding the outgoing coordinate description is a formal mathematical 
mechanism for breaking $T$ symmetry. However, there is no physical justification for this action\footnote{A conceptually
parallel situation is seen in classical electrodynamics where advanced wave solutions are discarded. However, in
that situation there is the physical observation of causality that justifies the action. Besides, quantum
electrodynamics eventually resolves the problem without breaking $T$ symmetry.}.

\section{Nonconservative Forces and Time Reversal}
For sources of gravity that are not black holes (say the Earth), $T$ symmetry violation is quite common. Meteorites
falling on the Earth do not return to orbit. This happens due to nonconservative forces like air friction and
ground impact. For black holes, one does not know of such forces. But they could be postulated. They would then
produce some kind of radiation to maintain overall energy conservation. Interestingly, this radiation should be 
expected to escape as it starts off being outwardly directed. However, in the absence of any known cause of such forces,
it is ``natural'' not to postulate them. Hence, maintaining $T$ symmetry would be more ``natural''.

\section{The Equation of Motion -- Classical Scattering}
Consider the classical scattering problem of a beam of particles incident on a black hole. Equations~\ref{eqmotion},
\ref{eqangmom} and~\ref{eqradial} 
provide solutions for such a problem. However, unlike typical scattering problems, this one can have very different
physical meanings for different observers. For observers moving along with individual particles of the beam,
the time coordinates are their individual proper times $\tau$ and
they see $T$ symmetric trajectories as long as no nonconservative forces are introduced at the singular point.
On the other hand, a distant stationary observer uses the coordinate time $t$. As a result, he/she does not see
complete trajectories for particles that enter the black hole. Hence, for this observer, the scattering process
may seem inelastic as some of the particles of the incident beam never reappear at a distant point. However, closer
scrutiny shows this to be sort of a misdirection. This is because, for the distant observer,
this is not even a complete scattering problem. Experimental observations for a typical scatterirng problem are
made once the system reaches a steady state. The distant observer never sees the system reach that steady state!
The particles falling into the black hole are never done with their falling. So, the question of inelasticity is moot.

In this light, it is interesting to address the problem of quantum scattering due to a black hole. This has 
been done elsewhere\cite{vach}.

\section{Computation of Time Reversal Symmetric Trajectories}
In order to maintain $T$ symmetry, particle trajectories that fall to the center need to be continued beyond that point.
Although, the singularity does not allow computation exactly at $r=0$, a physically meaningful analytical continuation
that skips over the singularity is possible. It requires energy and angular momentum conservation.
The process is somewhat different for radial ($h=0$) and non-radial ($h\neq 0$) trajectories. So, the two cases will
be treated separately in the following.

The ingoing part of a radial trajectory, at some angle $\phi=\phi_{0}$, is computed using equation~\ref{eqradial}
up to a point where $r=\epsilon\rightarrow 0$. Then, to continue the trajectory on the other side of the singularity as
the outgoing part of the trajectory, the following initial conditions are used.
\beq
r=\epsilon,\;\;\phi=\phi_{0}-180^{\circ},\;\;\frac{dr}{d\tau}=-\left(\frac{dr}{d\tau}\right)_{\epsilon},
\eeq
where $(dr/d\tau)_{\epsilon}$ is the value of $dr/d\tau$ at $r=\epsilon$ on the ingoing side of the trajectory. The second
condition is justified by the fact that the conserved angular momentum in this case is zero and hence the particle cannot
change its direction. The third condition is due to energy conservation as, for radial trajectories, the energy $e$ 
is expected to be a function of $r$ and $dr/d\tau$ as follows.
\beq
e\equiv e(r,(dr/d\tau)^{2}).
\eeq
Note that an explicit functional form for $e$ is avoided as that would bring forth a debate over controversial interpretations
of coordinates. 

For non-radial trajectories, $\phi$ is not a constant and equation~\ref{eqmotion} is used for computations. For 
trajectories that go through the center, once again, the ingoing part is computed up to a point where 
$r=\epsilon\rightarrow 0$. Then, to continue the 
trajectory on the other side of the singularity as the outgoing part of the trajectory, the following initial 
conditions are used.
\beq
r=\epsilon,\;\;\phi=\phi_{\epsilon}-180^{\circ},\;\;\frac{du}{d\phi}=-\left(\frac{du}{d\phi}\right)_{\epsilon},
\eeq
where $\phi_{\epsilon}$ and $(du/d\phi)_{\epsilon}$ are the values of $\phi$ and $du/d\phi$ at $r=\epsilon$ on the 
ingoing side of the trajectory. The third condition is due to conservation of energy $e$ and angular momentum $h$. 
This is because the energy $e$, in this case, is expected to be a function of $r$, $h$ and $du/d\phi$ as follows.
\beq
e\equiv e(r,h^{2},(du/d\phi)^{2}).
\eeq
The second initial condition, in this case, needs more detailed justification as $d\phi/d\tau\rightarrow\infty$ at $r=0$ for
$h\neq 0$ (equation~\ref{eqangmom}). This raises the possibility of
discontinuous changes in the direction of motion at the origin. Hence, in the following it is proved that the
direction of motion does indeed remain continuous at the origin as indicated by the second initial condition.

Although, $d\phi/d\tau\rightarrow\infty$ at the origin, that, by itself, cannot produce a discontinuous change in direction.
For changes in direction, we need to follow the behavior of $d\phi/dr$. To this end, we consider the following integral of equation~\ref{eqmotion}\cite{rindler,night}.
\beq
\left(\frac{du}{d\phi}\right)^{2}+u^{2}=E+\frac{r_{s}u}{h^{2}}+r_{s}u^{3},
\eeq
where $E$ is a constant. Then, for $r=\epsilon\rightarrow 0$,
\beq
\frac{du}{d\phi}=r_{s}^{1/2}u^{3/2}.
\eeq
Hence,
\beq
\frac{d\phi}{dr}=-r_{s}^{-1/2}r^{-1/2}.
\eeq
If $\Delta\phi$ is the change in $\phi$ for an $\epsilon$ change in $r$ at the origin, then
\beq
\lim_{\epsilon\rightarrow 0}\frac{\Delta\phi}{\epsilon}=\left.\frac{d\phi}{dr}\right|_{r=0}=
-\lim_{\epsilon\rightarrow 0}(r_{s}^{-1/2}\epsilon^{-1/2}).
\eeq
Hence,
\beq
\lim_{\epsilon\rightarrow 0}\Delta\phi=-\lim_{\epsilon\rightarrow 0}(\epsilon^{1/2}r_{s}^{-1/2})=0.
\eeq
This proves the continuity of the direction of motion at the origin.

\begin{figure}
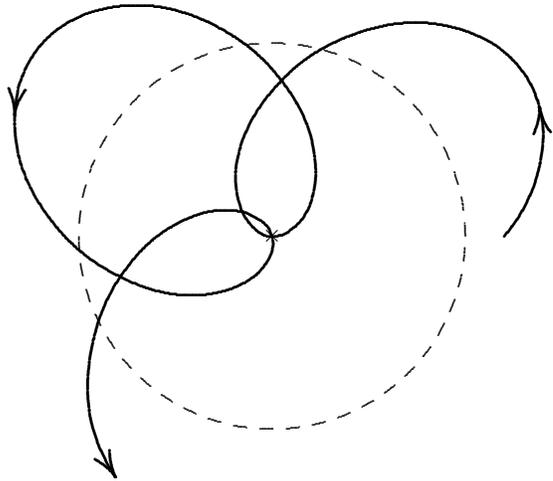

\beginpicture
\setcoordinatesystem units <.07mm,.07mm>
\setplotarea x from -600 to 600, y from -500 to 500
\setdashes
\circulararc 360 degrees from 367 0 center at 0 0
\setsolid
\setplotsymbol ({\rm .})
\linethickness=1pt
\setquadratic
\plot
441	0
495	90
515	194
489	293
418	367
319	402
213	400
118	368
41	318
-12	261
-47	204
-65	150
-69	103
-63	65
-51	36
-35	16
-20	5
-8	0
17	1
32	9
50	24
67	48
79	83
83	127
76	180
54	240
12	303
-50	364
-135	413
-238	438
-343	426
-431	372
-481	284
-486	179
-452	79
-392	-2
-320	-60
-248	-95
-182	-111
-125	-111
-78	-100
-43	-83
-19	-62
-5	-41
1	-22
2	-9
-6	15
-18	27
-38	39
-66	48
-103	49
-146	39
-194	16
-244	-23
-291	-82
-329	-161
-349	-257
-341	-361
-297	-458
/
\put {$\times$} at 0 0
\arrow <10pt> [.2,.67] from 508.1 244 to 508	245
\arrow <10pt> [.2,.67] from -488.9 232 to -489 231
\arrow <10pt> [.2,.67] from -297.6 -457 to -297 -458
\endpicture
\caption{Trajectory of a bound particle. \label{fig1}}
\end{figure}

\begin{figure}
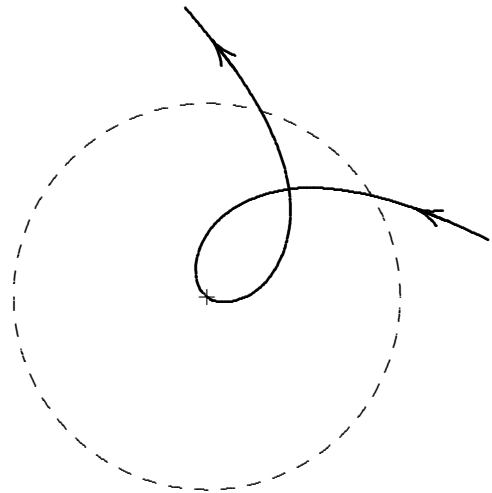

\beginpicture
\setcoordinatesystem units <.07mm,.07mm>
\setplotarea x from -600 to 600, y from -500 to 500
\setdashes
\circulararc 360 degrees from 367 0 center at 0 0
\setsolid
\setplotsymbol ({\rm .})
\linethickness=1pt
\setquadratic
\plot
534	108
460	142
396	167
340	185
289	197
243	204
201	207
162	205
128	199
97	190
69	178
45	164
25	148
9	131
-3	113
-12	96
-18	78
-21	62
-21	47
-20	34
-17	23
-13	15
-9	8
5	-4
11	-7
19	-9
28	-10
40	-10
53	-8
67	-3
82	3
97	14
112	28
125	45
137	66
147	90
154	117
158	147
158	179
154	215
145	252
131	292
112	335
86	381
53	432
10	487
-42	550
/
\put {$+$} at 0 0
\arrow <10pt> [.2,.67] from 460	142 to 396	167
\arrow <10pt> [.2,.67] from 53	432 to 10	487
\endpicture
\caption{Trajectory of a scattered particle. \label{fig2}}
\end{figure}

Analytical solutions for radial trajectories are straightforward\cite{night}. For non-radial trajectories,
equation~\ref{eqmotion} needs to be solved numerically. For this purpose, the value of $\epsilon$ must be
non-zero but small. It should be chosen to minimize error. Estimation 
of error is done by reducing $\epsilon$ by a factor of half and noting the resulting change in trajectory.
Some results of numerical computation using a fourth order Runge-Kutta method are shown in 
figures~\ref{fig1} and~\ref{fig2}. The dashed line represents
the event horizon. Time reversal symmetry is apparent in these computed trajectories.

\section{Conclusion}
It is seen that the equation of motion of a particle falling into a black hole is time reversal symmetric. Hence, it is
natural to expect particle trajectories around a black hole to also be time reversal symmetric. However, trajectories that
end at the central singularity are clearly not so. Here it is shown that no trajectory has to end at the central
singularity. Particles should be able to go through the center and continue on an outward path. This would maintain
their time reversal symmetry.

It is usually assumed that particles that cross the event horizon enter a different realm of time that cannot be
connected to an outside observer's time. If particle trajectories are time reversal symmetric in proper time, there
should also be particles emerging out of this other realm of time. So, real black holes should not be expected
to be dark objects. Of course, as seen by the outside observer, the particles that come out are not the same as the 
ones that go in. Particles that go in do so forever without actually reaching the horizon in finite time. 
On the other hand, particles that are seen to come out would never be seen to actually emerge from the event horizon 
as that would have happened at $t=-\infty$.

Stellar objects that are expected to look dark are the ones that are collapsing and, according
to the outside observer, will continue to do so for infinite time without actually becoming a black hole. Of course,
in the early stages of collapse, a star is expected to produce observable electromagnetic radiation. As the
collapse progresses, the resulting strong gravity red shifts all electromagnetic radiation to a degree that makes
them unobservable. Hence, paradoxically, dark stars should not be expected to be black holes while actual 
primordial black holes can appear to be not dark at all.

\end{document}